\documentstyle[11pt,newpasp,twoside,psfig]{article}
\def\edcomment#1{\iffalse\marginpar{\raggedright\sl#1\/}\else\relax\fi}
\reversemarginpar

\begin{document}
\title{The Iron Project and Non-LTE stellar modeling}
 \author{Sultana N. Nahar}
\affil{Dept of Astronomy, Ohio State University, Columbus, OH
43210, USA }
%\author{Ima Co-Author}
%\affil{The Name of My Institution, The Full Address of My Institution}

\begin{abstract}
Latest developments in theoretical computations since the international
Opacity Project (OP), under the new the Iron Project (IP) and 
extensions, are described for applications to a variety of objects such 
as stellar atmospheres, nebulae, and active galactic nuclei. The primary 
atomic processes are: electron impact excitation (EIE), photoionization, 
electron-ion recombination, and bound-bound transitions, all considered 
using the accurate and powerful R-matrix method including relativistic 
effects. As an extension of the OP and the IP, a self-consistent and 
unified theoretical treatment of photoionization and recombination has 
been developed. Both the radiative and the dielectronic recombination 
(RR and DR) processes are considered in a unified manner. Photoionization 
and recombination cross sections are computed with identical 
wavefunction expansions, thus ensuring self-consistency in an ab initio 
manner. The new unified results differ from the sum of previous results 
for RR and DR by up to a factor of 4 for the important but complex atomic 
systems such as Fe~I~-~V. The fundamental differences are due to quantum 
mechanical intereference and coupling effects neglected in simpler 
approximations that unphysically treat RR and DR separately, which can 
not be independently measured or observed. The electronic, 
web-interactive, database, TIPTOPBASE, to archive the OP/IP data in a 
readily accessible manner is also described. TIPTOPBASE would include 
electron-ion recombination data and new fine structure transition 
probabilities. Efficient codes developed by M.J. Seaton to calculate 
`customized' mixture opacities and radiative accelerations ('levitation') 
in stars will also be available.

\end{abstract}

\section{Introduction}

At densities and temperatures in stellar atmospheres, many atomic levels 
are excited under non-local thermal equilibrium (NLTE) conditions. 
NLTE models and other applications such as stellar opacities require 
large amount of accurate atomic parameters for collisional and radiative 
processes to describe radiation transfer, spectral analysis etc.
The collisional process is primarily electron impact excitation (EIE), 
while radiative processes are photoionization, electron-ion 
recombination, and bound-bound transitions. These four basic dominant 
atomic processes in the plasmas can be described as follows.

\noindent
i) Electron-impact excitation (EIE) of an ion $X^{+z}$ of charge z:

$$e + X^{+z} \rightarrow e' + X^{+z*}.$$

\noindent
ii) Radiative bound-bound transitions:

$$X^{+z} + h\nu \rightleftharpoons X^{+z*},$$

\noindent
iii) Photoionization (PI) by absorption of a photon:

$$X^{+z} + h\nu \rightleftharpoons X^{+z+1} + \epsilon.$$

\noindent
The inverse process of PI is electron-ion radiative recombination (RR).

\noindent
iv) Autoionization (AI) and dielectronic recombination (DR):

$$e + X^{+z} \rightarrow (X^{+z-1})^{**} \rightarrow \left\{ \begin{array}{ll}
e + X^{+z} & \mbox{AI} \\ X^{+z-1} + h\nu & \mbox{DR} \end{array}
\right. $$

\noindent
The inverse process of DR is photoionization via the intermediate doubly 
excited autoionizing states, i.e. resonances in atomic processes.
At prevailing densities and temperatures in stellar atmospheres, the 
role of metastable states and low-lying fine structure levels in 
photoionization and recombination of ions bears special emphasis.

Collisional and radiative atomic process have been studied in ab initio 
manner under the OP ({\it The Opacity Project} 1995, 1996), and the IP 
(Hummer et. al 1993). The close coupling 
R-matrix methodology employed under the OP and IP enables the 
computation of self-consistent sets of atomic parameters, thereby 
reducing uncertainties in applications involving different processes 
and approximations. Sample results obtained under the two projects are 
presented.

\section{Theory}

Atomic processes are treated in an ab initio manner in the close 
coupling (CC) approximation employing the R-matrix method (e.g. Burke 
\& Robb 1975, Seaton 1987, Berrington et al. 1987, Berrington et al. 
1995). The total wavefunction for a (N+1) electron system in the CC
approximation is described as:
\begin{equation}
\Psi_E(e+ion) = A \sum_i^N \chi_i(ion)\theta_i + \sum_{j} c_j \Phi_j(e+ion),
\end{equation}
\noindent
where in the first term $\chi_i$ is the target ion or core wavefunction 
in a specific state $S_iL_i\pi_i$ or level $J_i\pi_i$, $\theta_i$ is the 
wavefunction of the interacting (N+1)th electron in a channel labeled as
$S_iL_i(J_i)\pi_i \ k_{i}^{2}\ell_i(SL\pi~or~ \ J\pi)$, $k_{i}^{2}$ is 
the incident kinetic energy.  In the second term, $\Phi_j$ is the 
correlation functions of (e+ion) system that compensates the orthogonality 
condition and short range correlation interations. The complex resonant 
structures in photoionization, recombination, and in electron impact 
excitation are included through channel couplings. The target wavefunctions
are obtained from configuration interaction atomic structure 
calculations using code, such as, SUPERSTRUCTURE (Eissner et al. 1974).

The relativistic (N+1)-electron Hamiltonian for the N-electron target
ion and a free electron in the Breit-Pauli approximation, as adopted under
the IP, is
\begin{equation}
H_{N+1}^{\rm BP}=H^{NR}_{N+1}+H_{N+1}^{\rm mass} + H_{N+1}^{\rm Dar}
+ H_{N+1}^{\rm so},
\end{equation}
\noindent
where non-relativisitc Hamiltonian is
\begin{equation}
H^{NR}_{N+1} = \sum_{i=1}\sp{N+1}\left\{-\nabla_i\sp 2 - \frac{2Z}{r_i}
        + \sum_{j>i}\sp{N+1} \frac{2}{r_{ij}}\right\}.
\end{equation}
\noindent
The mass correction, Darwin, and spin-orbit interaction terms are:
\begin{equation}
H_{N+1}^{\rm mass} = -{\alpha^2\over 4}\sum_i{p_i^4}, ~
H_{N+1}^{\rm Dar} = {Z\alpha^2 \over 4}\sum_i{\nabla^2({1\over r_i})}, ~
H_{N+1}^{\rm so}= Z\alpha^2 \sum_i{1\over r_i^3}{\bf l_i.s_i}.
\end{equation}
Spin-orbit interaction splits the LS terms into 
fine-structure $J$-levels.

The set of ${SL\pi}$ are recoupled to obtain (e + ion) states  with
total $J\pi$, following the diagonalization of the (N+1)-electron
%Hamiltonian to solve $H^{BP}_{N+1}\mit\Psi_E = E\mit\Psi_E$.
Hamiltonian to solve 
\begin{equation}
H^{BP}_{N+1}\mit\Psi_E = E\mit\Psi_E.
\end{equation}
Substitution of the wavefunction expansion introduces a set of coupled
equations that are solved using the R-matrix approach. The continuun 
wavefunction, $\Psi_F$, describe the scattering process with the free 
electron interacting with the target at positive energies (E $>$ 0), 
while at {\it negative} total energies (E $<$ 0), the solutions correspond 
to pure bound states $\Psi_B$.

\subsection{Electron Impact Excitation}

Electron impact excitation is one of the primary processes for spectral 
formation in astrophysical and laboratory plasmas. The collision 
strength for transition by electron impact excitation from the initial 
state of the target ion $S_iL_i$ to the final state $S_jL_j$ is given by
\begin{equation}
\Omega(S_iL_i-S_jL_j) = {1\over 2}\sum_{SL\pi}\sum_{l_il_j}(2S+1)
(2L+1)|{\bf S}^{SL\pi}(S_iL_il_i - S_jL_jl_j)|^2,
\end{equation}
where ${\bf S}$ is the scattering matrix, and $S$, $L$ are the spin
multiplicity and total orbital angular momentum of the (e,ion) system. 
The effective collision strength or the Maxwellian averaged collision 
strength can be obtained as
\begin{equation}
\Upsilon(T)=\int_0^{\infty} \Omega_{ij}(\epsilon_j)e^{-\epsilon_j \over kT}
d(\epsilon_j/kT),
\end{equation}
and the excitation rate coefficient for transtion from level i 
$\longrightarrow$ j as, $q_{ij}(T)$ = (8.63$\times 10^{-6}/ g_iT^{1/2}) 
e^{-E_{ij}/kT} \Upsilon (T)$ in $~cm^3s^{-1}$, where $T$ is in K, 
$E_{ij}=E_j-E_i$, $E_i<E_j$ are in Rydbergs (1/kT = 157885/T), and $g_i$ 
is the statistical weight of i.

\subsection{Photoionization, Recombination, Transition Probabilities}

The transition matrix elements for radiative bound-bound excitation 
or de-excitation can be obtained using bound-state wavefunctions as 
$<\Psi_B || {\bf D} || \Psi_{B'}>$, and for photoionization and 
recombination using the bound and continuum wavefunctions as 
$<\Psi_B || {\bf D} || \Psi_{F}>$; 
{\bf D} is the dipole operator. In "length" form, ${\bf D}_L = 
\sum_i{r_i}$, and in "velocity" form, ${\bf D}_V = -2\sum_i{\Delta_i}$,
where the sum corresponds to number of electrons.

The transition matrix element with the dipole operator can be reduced
to the generalized line strength defined, in either length or velocity
form, as
\begin{equation}
S_{\rm L}= |<\Psi_j||{\bf D}_L||\Psi_i>|^2 =
 \left|\left\langle{\mit\Psi}_f
 \vert\sum_{j=1}^{N+1} r_j\vert
 {\mit\Psi}_i\right\rangle\right|^2 \label{eq:SLe},
\end{equation}
\begin{equation}
S_{\rm V}=E_{ij}^{-2}|<\Psi_j||{\bf D}_V||\Psi_i>|^2 = \omega^{-2}
 \left|\left\langle{\mit\Psi}_f
 \vert\sum_{j=1}^{N+1} \frac{\partial}{\partial r_j}\vert
 {\mit\Psi}_i\right\rangle\right|^2. \label{eq:SVe}
\end{equation}
where $\omega$ is the incident photon energy in Rydberg units, and
$\mit\Psi_i$ and $\mit\Psi_f$ are the initial and final state wave functions.

The oscillator strength $f_{ij}$ and the transition probability $A_{ji}$
for the bound-bound transition are obtained in atomic units (a.u.) as
\begin{equation}
f_{ij} = {E_{ji}\over {3g_i}}S,
~~A_{ji}(a.u.) = {1\over 2}\alpha^3{g_i\over g_j}E_{ji}^2f_{ij},
%~~~~~ \tau_j
%= {1\over \sum_i A_{ji}(s^{-1})},
\end{equation}
where $E_{ji}$ is the transition energy, $\alpha$ is the fine structure
constant, and $g_i$, $g_j$ are the statistical weights. The lifetime 
of a level $j$ decaying to all lower levels $i$, is $\tau_j = 
({\sum_i A_{ji}(s^{-1})})^{-1}$,  $A_{ji}(s^{-1}) = {A_{ji}(a.u.)/\tau_0}$,
$\tau_0 = 2.4191\times 10^{-17}$ is the atomic unit of time.

The photoionization cross section ($\sigma_{PI}$) is proportional to
the generalized line strength ($S$),
\begin{equation}
\sigma_{PI} = {4\pi \over 3c}{1\over g_i}\omega S.
\end{equation}

The unified electron-ion recombination method
(Nahar \& Pradhan 1994,1995) considers the infinite number of 
recombined states, and incorporates non-resonant and resonant
(radiative and dielectronic) recombinations RR and DR.
The recombined states are divided into two groups. The 
contributions from states with $n \leq$ 10 (group A) are obtained from 
$\sigma_{PI}$ using principle of detailed balance, while the 
contributions from states with 10 $< n \leq ~\infty$ (group B), which
are dominated by narrow dense resonances, are obtained from an extension 
of the DR theory of Bell \& Seaton (1985, Nahar \& Pradhan 1994).

The recombination cross section, $\sigma_{RC}$, is related to
$\sigma_{PI}$ through the principle of detailed balance,
\begin{equation}
\sigma_{RC} = \sigma_{PI}{g_i\over g_j}{h^2\omega^2\over 4\pi^2m^2c^2v^2}.
\end{equation}
The recombination rate coefficient, $\alpha_{RC}$, is obtained from 
Maxwellian average over $\sigma_{RC}$ as
\begin{equation}
\alpha_{RC}(T) = \int_0^{\infty}{vf(v)\sigma_{RC}dv},
\end{equation}
where $f(v)$ is the Maxwellian velocity distribution function. However,
the total $\alpha_{RC}$ is obtained from the summed contributions from 
infinite number of recombined states.

The DR cross sections of the high-n states is, $\sigma_{DR}(k^2;i,j) =
g_ik^2 \Omega_{DR}(k^2;i,j)$.
where the DR collision strength is
\begin{equation}
\Omega(DR) = \sum_{SL\pi} \sum_n (1/2)(2S+1)(2L+1) P^{SL\pi}_{n}.
\end{equation}
$P^{SL\pi}_n$ is the DR probability in entrance channel $n$ (Nahar
\& Pradhan 1994).

\subsection{Ionization fractions in plasma equilibrium}

The main application of self-consistent sets of atomic data for 
photoionization and electon-ion recombination is in determining ionization 
fractions in photoionization or collisional (coronal) equilibrium in 
astrophysical plasmas. Coronal equilibrium corresponds to the balance 
between electron impact ionization and the electron-ion recombination
\begin{equation}
N(z-1)S(z-1) = N(z)\alpha_{RC}(z)
\end{equation} 
where $S(z-1)$ is total electron impact ionization rate coefficient 
which are in general available from various experiments. 
Photoionization equilibrim in the presence of dominant radiative source 
is balance photoionization and electron-ion recombination
\begin{equation}
 N(z)\int_{\nu_0}^\infty {4\pi J_{\nu}\over h\nu}\sigma_{PI}(z,\nu)d\nu
 = N_eN(z+1)\alpha_{RC}(z,T_e),
\end{equation}
where $J_{\nu}$ is photoionizing radiation flux, $\nu_o$ is ionization 
potential of the ion, and $\sigma_{PI}$ refers to photoionization 
cross sections. Use of self-consistent data for $\sigma_{PI}$ on the 
left-hand-side, and $\alpha_{RC}$ on the right-hand-side, should yield
accurate ionization balance.

\section{Results from the Iron Project}

The collisional strengths and most of the radiative data are reported in 
the "Atomic data from the IRON Project" series in Astronomy and 
Astrophysics journal (and in its previous Supplements).  

Sample results are discussed for each atomic process below: Maxwellian
averaged excitation collision strengths $\Upsilon(T)$, 
photoionization cross sections $\sigma_{PI}$, recombination rate 
coefficients $\alpha_R(T)$, and $f$-values.

\subsection{Electron Impact Excitation}

One main aim of the IP is to compute collisional data for the iron-peak
elements in various ionization stages.  Collision strengths and rates 
for electron impact excitation (EIE) of all iron ions, Fe I - Fe XXVI, 
and other ions have been obtained. For example, Fe XVII, important in 
EUV and X-ray astronomy, has been recently studied with a large-scale 
BPRM calculation (Chen \& Pradhan 2002) using a 89-level wavefunction 
expansion including up to $n$ = 4 levels. Their results show extensive 
resonance structures in the collision strengths that considerably enhance 
the rate coefficients (Fig. 1a). The corresponding line ratios are in 
good agreement with observations (Fig. 1b); filled squares are ratios 
obtained using previous cross sections that differ from observations at 
low temperatures. This is of considerable importance in photoionized 
{\sc x}-ray plasmas that have temperatures of maximum abundance much 
lower than that in coronal equilibrium T$_m \sim  4 \times 10^6$ K for 
Fe XVII, as marked.

\begin{figure} %
\vspace*{-0.7cm}
\psfig{figure=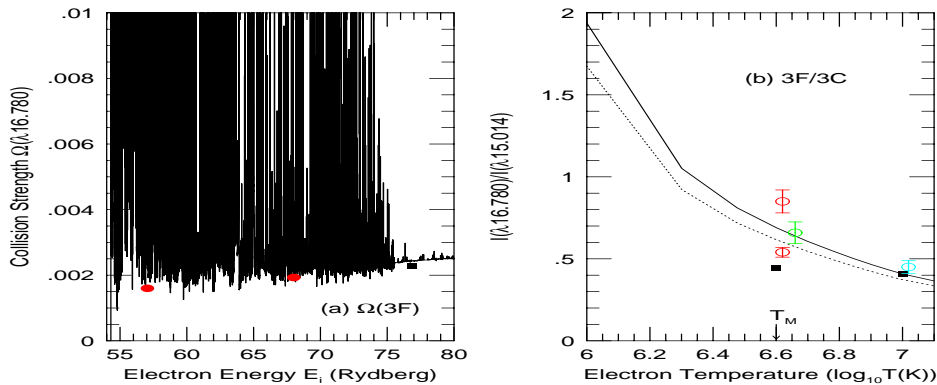,height=8.0cm,width=11.0cm}
\vspace*{-1.35cm}
\caption{(a) BPRM collision strength $\Omega$ for the forbidden
3F line; filled circles and square are non-resonant distorted wave 
calculations; (b): forbidden
to resonance line ratio 3F/3C vs.~T from a 89-level C-R model. 
The electron densities for
solid-line and dot-line curves are 10$^{13}$ and 10$^9$ cm$^{-3}$ 
respectively. The 4 open circles with error bars are observed and 
experimental values; filled squares are values using distorted wave
cross sections. } 
\end{figure}

Reviews and extensive compilations of available theoretical data sources 
for EIE collision strengths can be found in Pradhan \& Galagher (1992), 
Pradhan \& Peng (1995) and Pradhan \& Zhang (2001). A table of recommended 
data for effective collision strengths and A-values for ions from these 
references is available on-line from
www.astronomy.ohio-state.edu/$\sim$pradhan

\subsection{Photoionization and Recombination}

The CC approximation, utilising the powerful R-matrix method, yields in 
an ab initio manner: (A) self-consistent photoionization and recombination 
cross sections using identical wavefunction expansions for both processes 
over all energies, (B) unified e-ion recombination (RR and DR) rates at 
all temperatures of interest, and (C) level-specific recombination rates 
for a large number of atomic levels. In contrast to simple approximations 
that artificially treat RR and DR separately, with different methods and
over limited energy and temperature ranges, the present method accounts
for e-ion recombination as it occurs in nature. 

\subsubsection{Photoionization}

The CC approximation enables the calculation of photoionization cross 
sections $\sigma_{PI}$ for the ground and large number of excited bound 
states; both OP and IP typically consider all states up to $n\leq$ 10. 
The cross
sections include delineated autoionizing resonances upto $\nu \leq$ 10,
for each Rydberg series belonging to the various core thresholds.
The resonances can enhance the background cross sections considerably. 
Fig.~2 shows the ground state photoionization cross sections of 
Fe~I to Fe~V (Bautista, Nahar, \& Pradhan 1994-1997) dominated by 
extensive resonances. The enhancement in the backgound is up to three 
orders of magnitude for Fe I, over an order or magnitude for Fe II, and 
$\sim$ 50\% for Fe III.

\begin{figure} %
\vspace*{-2.5cm}
\psfig{figure=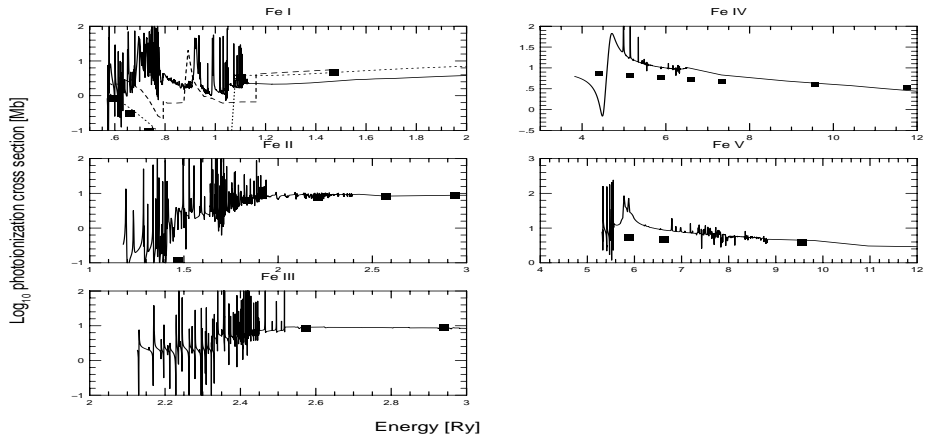,height=9.0cm,width=11.0cm}
\vspace*{-1.0cm}
\caption{Photoionization cross sections, $\sigma_{PI}$, of the ground
state of Fe I - Fe V, show large enhancements (the Y-axis is on a
Log-scale) compared with previous works without resonances
(e.g. Reilman and Manson 1979, filled squares; and Verner et al 1993).}
\end{figure}

Channel couplings show important features that are nonexistent in 
simpler (e.g. central-field or screened-hydrogenic) approximations. 
One important feature is the photo-excitation-of-core (PEC) resonances 
in the excited state $\sigma_{PI}$. Fig. 3 shows $\sigma_{PI}$ of excited 
series of states, $3d^5ns(^7S)$ with $5\leq n \geq$ 11, of Fe III.
The background cross section falls monotonically for each state until 
$\sim$ 1.73 Ry, where a strong and wide PEC resonance (pointed by the 
arrow) enhances the background considerably. The energy corresponds to 
the excited core threshold $3d^44p(^6P^o)$ state of Fe IV. The photon 
at this energy is absorbed by the core in the dipole allowed transition 
from the ground state $3d^5(^6S)~\rightarrow ~3d^44p(^6P^o)$, while the 
valence electron remains a `spectator'. The electron is ejected when the 
core drops to ground state (PEC is the inverse process of DR). This 
enhancement contradicts the usual assumption of hydrogenic behavior of 
excited states, and can affect both the photoionization and recombination 
rates at high temeperatures.  

\begin{figure} %
\vspace*{0.2cm}
\psfig{figure=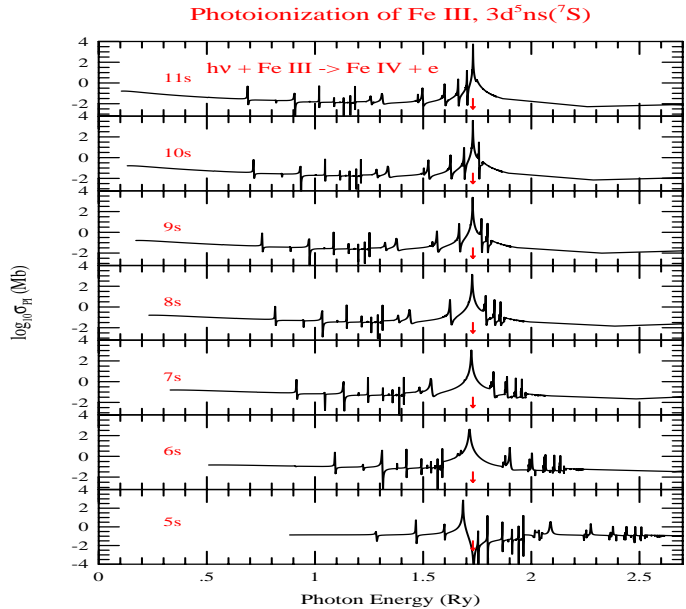,height=9.0cm,width=11.0cm}
\vspace*{-0.5cm}
\caption{Photoionization cross sections, $\sigma_{PI}$, of the excited
states of states, $3d^5np(^7P^o)$, of Fe III illustrating large
PEC resonance.}
\end{figure}

All $\sigma_{PI}$ obtained under the OP, and for some complex ions, $LS$ 
coupling approximation has been employed. However, relativistic BPRM 
approximation is now implemented for highly charged ions, such as Fe XXIV, 
Fe XXV, C IV, C V (Nahar, Pradhan, Zhang 2000, 2001) etc. Some large 
atomic systems, such as Fe XVII (Zhang, Nahar \& Pradhan 2001), are under 
investigation including relativistic fine structure.

\subsubsection{Electron-ion Recombination}

Fig. 4 presents the total unified recombination rate coefficient 
$\alpha_R(T)$ for iron ions, Fe I - Fe V (solid, Nahar, Bautista, \& 
Pradhan 1996-1999). The general features of the unified rate over a wide 
temperature ($T$) range are as follows: At very low $T$, when electrons 
are not energetic enough to form doubly excited autoionizing states, 
recombination is dominated by RR. The rate decreases with $T$ until, at 
high $T$, when it rises again due to dominance by DR. Comparison of the 
unified rates with the {\it sum} of the earlier RR rates
from central field calculations, and the high temperature DR rates
mainly from the Burgess general formula, show either an underestimate
or overestimate by several factors at temperatures of maximum abundance
of Fe~I - V (dashed lines in lower panels in Fig. 4). The large 
differences between the unfiied rate and the sum of the earlier (RR+DR) 
results illustrates the fundamental inaccuracy of separate treatments 
of RR and DR. Resonances are inextricably related to the coupling within 
the wavefunction expansion, and may not be separated from the non-resonant
background, as seen in the photoionization cross sections in Fig. 3.
The inclusion of this quantum mechanical intereference in the R-matrix
method is the basic advance enabled by the unified method.

Recently Gorczyca et.al. (2002) discuss certain effects of marginal
importance, such as radiation damping of some resonances, radiative 
decay beween autoionizing states, and forbidden transitions in the core 
(DR is dominated by dipole allowed transitions). They selectively 
consider small energy ranges, and/or a few resonances, in one or two 
highly charged ions where a separation between RR and DR may not be too 
inaccurate. Although Gorczyca et. al. describe the neglect of these 
effects as 'shortcomings of the R-matrix method', the overall effect on 
total recombination rates of practical astrophysical importance does 
not exceed about 10\% - well within the accuracy of the R-matrix method - 
even for their example of Fe~XVIII. In any event, these effects 
(particularly radiation damping) are, or may be, considered in our 
formulation if necessary. We note that other effects, such as external 
field ionization of high-$n$ 
Rydberg levels, are likely to be much more important. We emphasize that
the unified R-matrix method is the only method generally capable of
computing unified recombination rates, and
consistent with scattering and photoionization calculations, for
all atomic systems.

\begin{figure} %
%\vspace*{-0.7cm}
\hspace*{-1.5cm}
\psfig{figure=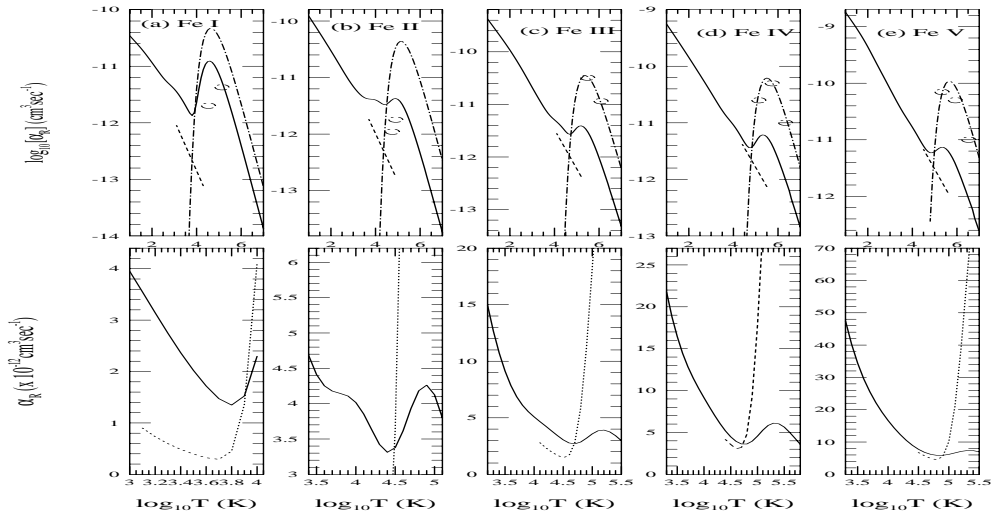,height=8.0cm,width=16.0cm}
%\vspace*{-0.7cm}
\caption{Unified total recombination rate coefficients ($\alpha_R(T)$ 
(solid), compared with the previous RR and DR calculations, in upper 
panels.  Comparison with the sum of RR+DR rates computed separately in
previous works is shown in lower panels on an expanded scale close to 
temperatures of maximum ionic abundance; For example, the unified rate
for Fe~I is higher than previous (RR+DR) rates by up to a factor of 4
for T $< 10^4$ K.}
\end{figure}

The unified rates are valid over a wide range of temperatures for all 
practical purposes, in contrast to RR and DR rates obtained using 
different approximations in different temperature ranges.  Following is 
the list of over 45 atoms and ions for which self-consistent sets of 
$\sigma_{PI}$ and $\alpha_R(T)$ have so far been obtained (reported in 
Nahar and Pradhan 1997, and subsequent publications in ApJS, and in 
others and on $www.astronomy.ohio-state.edu/\sim pradhan$):

\noindent
Carbon: C I, C II, C III, C IV, C V, C VI \\
Nitrogen: N I, N II, N II, N IV, N V, N VI, N VI \\
Oxygen: O I, O II, O III, O IV, O V, O VI, O VII, O VII \\
Silicon: Si I, Si II, Si IX \\
Sulfur: S II, S III, S XI \\
Iron: Fe I, Fe II, Fe III, Fe IV, Fe V, Fe XIII, Fe XVII,
Fe XXI, Fe XXIV, Fe XXV, Fe XXVI \\
C-like: F IV, Ne V, Na VI, Mg VII, Al VIII, Ar XIII, Ca XV \\
Other ions: Ar V, Ca VII, Ni II

\noindent

The complete data include state specific photoionization cross sections
and recombination rates for hundreds of bound levels with n$\le$ 10 for 
each ion.

\subsection{Transition probabilities of atoms and ions}

Under the IP transition probabilities are evaluated in the BPRM 
approximation including relativistic fine structure, in contrast 
OP work for dipole allowed $LS$ multiplets. Both the dipole allowed and 
spin-forbiden E1 intercombination transitions are obtained. Owing to fine 
structure, the sets of data for $f$- and $A$-values are considerably 
larger. For example, the large scale calculations for Fe V transition
probabilities resulted in about 1.5$\times 10^6$ transitions among
3865 fine structure levels (Nahar et al. 2000).

The fobidden electric quadrupole (E2) and octupole (E3), magnetic 
dipole (M1) and quadrupole (M2), transitions are also being evaluated 
in atomic structure calculations using SUPERSTRUCTURE (Eissner et al. 
1974) and other codes.

The BPRM fine structure energy levels are analysed using quantum defects 
and channel weights to obtain complete spectroscopic identifications, 
$(C_t \ S_t \ L_t \ J_t$ $\pi_t n\ell)SL J \ \pi$ where
$(C_t \ S_t \ L_t \ J_t~\pi_t)$ denotes the N-electron core configuration 
$C_t$, spin and orbital angular momenta $S_tL_t$, and parity$\pi_t$; 
$nl$ the outer or valence electron, $SL$ are the total spin and orbital 
angular momenta, and $J\pi$ are the total angular momentum and parity
of the (N+1)-electron system. 

Accurate fine structure transition probabilities have been obtained for 
a number of ions including: \\
Fe ions: Fe V, Fe XVII, Fe XXI, Fe XXIII, Fe XXIV, Fe XXV  \\
Li-like: C IV, N V, O VI, F VII, Ne VIII, Na IX, Mg X, Al XI, Si XII,
S XIV, Ar XVI, Ca XVIII, Ti XX, Cr XXII, Ni XXVI \\
He-like: C IV, N V, O VI, F VII, Ne VIII, Na IX, Mg X, Al XI, Si XII,
S XIV, Ar XVI, Ca XVIII, Ti XX, Cr XXII, Ni XXVI \\
Other ions: C II, C III, O IV, S II, Ar XIII, Na III, Cl-like ions

The latest BPRM work on transitions probabilities of Fe XVII 
has resulted in 2.6$\times 10^4$ dipole allowed and intercombination
E1 transitions. They correspond to 490 fine structure levels of 1/2 $\leq 
J \leq$ 17/2 of even and odd parities with $\leq n\leq $ 10 \& 0 
$\leq l\leq $ 9, $0 \leq L \leq 8$, and core $2s^22p(^2P^o_{3/2,1/2})$, 
$2s2p^6(^2S_{1/2})$. Oscillator strengths for about 360 electric 
quadrupole and magnetic dipole and a large number of E3 and M2 
transitions have also been obtained. A sample table of transitions
in Fe XVII with spectroscopic identifications is presented in Table I.

\begin{table}
\noindent{Table I: Transition probabilities of Fe XVII. g=2J+1. \\}
\normalsize
\begin{tabular}{llllrrrcl}
\hline
\noalign{\smallskip}
\hline
$C_i$ & $C_j$  & $T_i$ & $T_j$ & $g_i$ & $g_j$ & $E_{ij}$ &
$f$ & $A$ \\
 & & & & & & $(\AA)$ & & $(s^{-1})$ \\
 \noalign{\smallskip}
\hline
 \noalign{\smallskip}
 $2s22p6      $ & $-~2s22p53s    $ & $^1S^e$ & $^3P^o$ &  1 &  3 &     17.1 &  1.223E-01 &  9.35E+11 \\
 & & & & & & & & \\
 $2s22p6      $ & $-~2s22p53s    $ & $^1S^e$ & $^1P^o$ &  1 &  3 &     16.8 &  1.008E-01 &  7.96E+11 \\
 & & & & & & & & \\
 $2s22p6      $ & $-~2s22p54s    $ & $^1S^e$ & $^3P^o$ &  1: 1 &  3: 8 &     12.7 &  2.286E-02 &  3.16E+11 \\
 & & & & & & & & \\
 $2s22p6      $ & $-~2s22p54s    $ & $^1S^e$ & $^1P^o$ &  1: 1 &  3: 9 &     12.5 &  1.758E-02 &  2.49E+11 \\
 & & & & & & & & \\
 $2s22p6      $ & $-~2s22p55s    $ & $^1S^e$ & $^3P^o$ &  1: 1 &  3:13 &     11.4 &  1.003E-02 &  1.71E+11 \\
 & & & & & & & & \\
 $2s22p6      $ & $-~2s22p55s    $ & $^1S^e$ & $^1P^o$ &  1: 1 &  3:14 &     11.3 &  1.219E-02 &  2.13E+11 \\
 & & & & & & & & \\
 $2s22p6      $ & $-~2s22p53d    $ & $^1S^e$ & $^3P^o$ &  1 &  3 &     15.4 &  8.136E-03 &  7.58E+10 \\
 & & & & & & & & \\
 $2s22p6      $ & $-~2s22p53d    $ & $^1S^e$ & $^3D^o$ &  1 &  3 &     15.3 &  6.208E-01 &  5.93E+12 \\
 & & & & & & & & \\
 $2s22p6      $ & $-~2s22p53d    $ & $^1S^e$ & $^1P^o$ &  1 &  3 &     15.0 &  2.314E+00 &  2.28E+13 \\
 & & & & & & & & \\
 $2s22p6      $ & $-~2s22p54d    $ & $^1S^e$ & $^3P^o$ &  1: 1 &  3:10 &     12.3 &  3.281E-03 &  4.81E+10 \\
 & & & & & & & & \\
 $2s22p6      $ & $-~2s22p54d    $ & $^1S^e$ & $^3D^o$ &  1: 1 &  3:11 &     12.3 &  3.594E-01 &  5.31E+12 \\
 & & & & & & & & \\
 $2s22p6      $ & $-~2s22p54d    $ & $^1S^e$ & $^1P^o$ &  1: 1 &  3:12 &     12.1 &  3.987E-01 &  6.03E+12 \\
 & & & & & & & & \\
 $2s22p6      $ & $-~2s2p63p     $ & $^1S^e$ & $^3P^o$ &  1: 1 &  3: 6 &     13.9 &  3.501E-02 &  4.03E+11 \\
 & & & & & & & & \\
 $2s22p6      $ & $-~2s2p63p     $ & $^1S^e$ & $^1P^o$ &  1: 1 &  3: 7 &     13.8 &  2.835E-01 &  3.30E+12 \\
 & & & & & & & & \\
 $2s22p6      $ & $-~2s2p64p     $ & $^1S^e$ & $^3P^o$ &  1: 1 &  3:18 &     11.0 &  1.073E-02 &  1.96E+11 \\
 & & & & & & & & \\
 $2s22p6      $ & $-~2s2p64p     $ & $^1S^e$ & $^1P^o$ &  1: 1 &  3:19 &     11.0 &  9.190E-02 &  1.68E+12 \\
 & & & & & & & & \\
 $2s22p53s    $ & $-~2s22p53p    $ & $^3P^o$ & $^3P^e$ &  3 &  1 &    296.0 &  3.354E-02 &  7.66E+09 \\
 $2s22p53s    $ & $-~2s22p53p    $ & $^3P^o$ & $^3P^e$ &  3 &  3 &    262.7 &  5.893E-05 &  5.70E+06 \\
 $2s22p53s    $ & $-~2s22p53p    $ & $^3P^o$ & $^3P^e$ &  5 &  3 &    252.5 &  4.985E-03 &  8.69E+08 \\
 $2s22p53s    $ & $-~2s22p53p    $ & $^3P^o$ & $^3P^e$ &  3 &  5 &    340.4 &  9.075E-02 &  3.13E+09 \\
 $2s22p53s    $ & $-~2s22p53p    $ & $^3P^o$ & $^3P^e$ &  5 &  5 &    323.5 &  6.913E-02 &  4.41E+09 \\
 \multicolumn{2}{l}{$LS$} & $^3P^o$ & $^3P^e$ &  9 &  9 &  &  8.959E-02 &  6.71E+09 \\
 & & & & & & & & \\
 $2s22p63s    $ & $-~2s22p53p    $ & $^1P^o$ & $^3P^e$ &  3 &  1 &    413.8 &  9.557E-03 &  1.12E+09 \\
 $2s22p63s    $ & $-~2s22p53p    $ & $^1P^o$ & $^3P^e$ &  3 &  3 &    351.6 &  4.162E-02 &  2.25E+09 \\
 $2s22p63s    $ & $-~2s22p53p    $ & $^1P^o$ & $^3P^e$ &  3 &  5 &    506.3 &  1.464E-03 &  2.29E+07 \\
\noalign{\smallskip}
\hline
\end{tabular}
\end{table}

\subsection{Ionization structure of a planetary nebula}

 Fig. 5 shows the ionization structure of Fe ions in a planetary nebula 
under typical conditions: Solid curves are ionic fractions from new cross
sections and recombination rates; the dotted and dashed curves are
from previous data. A large discrepancy is found for Fe V and Fe VI ionic
fractions, by a factor of two at T$_{eff}$ = 100,000 K.

\begin{figure} %
\vspace*{-1.5cm}
\psfig{figure=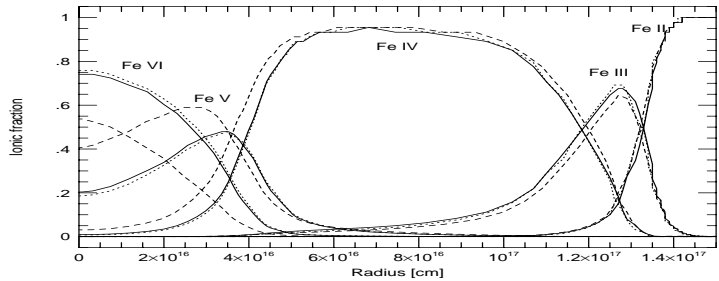,height=9.0cm,width=12.0cm}
\vspace*{-0.7cm}
\caption{Ionization fractions of Fe ions in a typical planetary nebula
obtained using the new data (solid), compared with those obtained using
previous calculations.}
\end{figure}

\section{TIPTOPBASE: atomic radiative and collisional data}

The current number of the series publication is 51. The IP website 
address is $http://www.usm.uni-muenchen.de/people/ip/iron-project.html$. 
The IP and related works can also be found at
$http://www.astronomy.ohio-state.edu/\sim pradhan$.
The extensive amount of atomic data and opacities obtained under the OP 
and the IP are available electronically through the existing database, 
TOPbase and through its planned extension 
TIPTOPBASE (C. Mendoza and the OP/IP team).

The radiative atomic and opacity data obtained under the OP are accessible 
via TOPbase from two websites:
($http://heasarc.gsfc.nasa.gov$ at Goddard, NASA and
$http://vizier.u-strasbg.fr/OP.html$ at CDS). TOPbase 
data, for ions with Z = 1 - 14, 16, 18, 20, 26, are: \\
(i) Photoionization cross sections of bound $LS$ terms, \\
(ii) Transition probabilities and $f$-values, \\
(ii) Energy levels, EQN ($\nu$), radiative lifetimes, \\
(iv) Monochromatic and Rosseland mean opacities 

The new database, TIPTOPbase (under development, C. Mendoza \& the OP/IP
team) will have collisional as well as radiative data, including

(i) All radiative data from TOPbase \\
(ii) Electron impact excitation collision strengths and rate coefficients
for all iron ions, Fe I - Fe XXVI, Ni~III, and other ions, \\
(iii) $A$-values for transitions in the target ion \\
(iv) Inner-shell photoionization ("tail") cross sections \\
(v) Radiative data for new elements through all ionization stages: P, 
Cl, K, and Ni II, Ni III, \\
(vi) More extended set of radiative data from repeated CC calculations \\
(vii) Radiative data ($\sigma_{PI}$ and $f$-values) for fine structure 
levels including relativistic effects \\
(viii) Total and level specific unified (RR+DR) recombination rate 
coefficients \\
(ix) Inner-shell radiative data for Fe ions - "PLUS" data \\
(x) On-line computational of opacities and radiative accelerations for 
user-specified mixtures of elements (``customized" opacities and 
radiative forces).

\subsection{TIPTOPbase: On-line computation}

Seaton has developed efficient codes for on-line computation of 
`customized' mixture opacities and radiative accelerations ('levitation') 
in stars.

i) Stellar opacities, Radiative forces: On-line interactive facility to 
compute mean and monochromatic opacities, and radiative forces, as 
function of temperature, density, and user-specified mixture of elements. 

1) Monochromatic opacities $\kappa_{\nu}$ for 17 elements: H, He, C, N, O, 
Ne, Na, Mg, Al, Si, S, Ar, Ca, Cr, Mn, Fe and Ni (Seaton et al 1994) as 
function of log($u = h\nu/kT$) at a mesh of (T,N$_e$) are tabulated.

(2) Tables of Rosseland mean $\kappa_R(T,\rho)$, with standard solar 
and non-solar, abundances where $\rho$ is the mass density (g/cc). 
$$ \frac {1}{\kappa_R} = \frac{\int_0^{\infty} \frac{1}
{\kappa_{\nu}} g(u)du} {\int_0^{\infty} g(u)du},
~~~g(u) = \frac{15}{4\pi^4} u^4e^{-u}(1-e^{-u})^{-2}, $$
where g(u) is the Planck weighting function have been calculated for 
different chemical compositions H (X),He (Y) and metals (Z), such that 
X+Y+Z = 1.

Partial derivatives of Rosseland means, $\left( \frac{\partial 
log(\kappa_R)}{\partial log(T)} \right)_{\rho}, \left( \frac{\partial 
log(\kappa_R)}{\partial log(\rho)} \right)_T , $ required for stellar 
structure and pulsation studies are also calculated. 

ii) Radiative accelarations:

Radiative acceleration, g$_{rad}$(k) for an element k is (Seaton 1997),
$$ g_{rad}(k) = (1/c) \int \sigma_{\nu} (k) F_{\nu} d\nu / M(k), $$
where M(k) is the mass of atom k and the flux F = $ \int F_{\nu} d\nu$
is related to the Rosseland mean $\kappa_R$ as $F_{\nu} = 
(\frac{\kappa_R}{\kappa_{\nu}}) (h/kT) g(u). $

\noindent
Radiative diffusion can lead to large changes in abundances of individual
elements in a star as gravitational forces are counteracted by radiative
levitation. For diffusion, one requires g$_{rad}$(k) for all depths in
the star, as a function of the abundance of element k. Interpolation
codes will be provided in TIPTOPbase to obtain radiative accelerations
for user-specified mixture and mesh of temperature and density.

\section{Conclusion}

The current status of large-scale ab initio close coupling R-matrix
calculations for radiative and collisional processes is reported.
The Iron Project Breit Pauli R-matrix radiative calculations include
large numbers of dipole allowed and intercombination transitions.
Self-consistent sets of atomic data for photoionization
and unified (electron-ion) recombination (including RR and DR)
are obtained, and should yield more accurate photoionization models.
Work is in progress for heavy ions of the iron group elements.

\acknowledgements{Partial supports by the U.S.  National Science 
Foundation and NASA are acknowledged.}

\end{document}